\documentclass[aps,pra,twocolumn,noshowpacs,noshowkeys,amsmath,amssymb,authoryear,nofootinbib]{revtex4}
\usepackage{graphicx}
\pdfoutput=1
\usepackage{bm}
\usepackage [autostyle, english = american]{csquotes}
\MakeOuterQuote{"}
\usepackage{dcolumn}
\usepackage{xcolor}
\usepackage[colorlinks,citecolor=blue,urlcolor=blue,bookmarks=false,hypertexnames=true,linkcolor=blue]{hyperref} 

\def\noi{\noindent}
\def\bc{\begin{center}}
\def\ec{\end{center}}
\newcommand{\bea}{\begin{equation}}
\newcommand{\eea}{\end{equation}\noi}
\newcommand{\ber}{\begin{eqnarray}}
\newcommand{\eer}{\end{eqnarray}\noi}
\begin{document}
\title{Applying the Nash Bargaining Solution for a Reasonable Royalty }

\author{David M. Kryskowski (Student)$^{1}$}\email{davidkryskowski@gmail.com}
\author{David Kryskowski$^{2}$}
\affiliation{$^{1}${Wayne State University, Detroit, Michigan, USA}\\
$^{2}$UD Holdings, 2214 Yorktown Dr., Ann Arbor, Michigan, 48105, USA}

\date{\today}

\begin{abstract}
There has been limited success applying the Nash Bargaining Solution (NBS) in assigning intellectual property damages due to the difficulty of relating it to the specific facts of the case. Because of this, parties are not taking advantage of \textit{Georgia-Pacific} factor fifteen. This paper intends to bring clarity to the NBS so it can be applied to the facts of a case. This paper normalizes the NBS and provides a methodology for determining the bargaining weight in Nash's solution. Several examples demonstrate this normalized form, and a nomograph is added for computational ease. 
\end{abstract}





\maketitle

\section{Introduction}

In U.S. patent litigation, there are two predominant ways to compensate a licensor when a firm infringes on its intellectual property. One way is to calculate the profit that was lost due to the infringement. The other way is to designate a reasonable royalty. A reasonable royalty is defined as a royalty assigned to the licensor to use its intellectual property by the licensee that is fair to both parties \citep{linck1994patent}.

Assigning a reasonable royalty is especially difficult in a dispute situation because of the difficulty of an arbiter or Court to attribute a royalty that is perceived as fair for both parties. A famous District Court case \textit{Georgia-Pacific vs. United States Plywood Corp}\footnote{\textit{Georgia-Pacific Corp. v. U.S. Plywood Corp.}, 318 F. Supp. 1116,
1120 (S.D.N.Y.1970), mod. and aff'd, 446 F.2d 295 (2d Cir. 1971), cert. denied, 404 U.S. 870 (1971).} demonstrated the complexity of assigning a reasonable royalty in litigation involving patents. As a result of the case, the District Court established fifteen guidelines for determining a reasonable royalty. However, guideline fifteen allowed for the use of a hypothetical license negotiation when the infringement began. This guideline implies that the NBS can be used as a justification for assigning a reasonable royalty.

In recent court cases, some judges have steered clear from using the NBS because parties often do not apply it to the specific facts of the case\footnote{Notable cases include: \textit{VirnetX, Inc. v. Cisco Systems, Inc.}, 767 F.3d 1308 (Fed. Cir. 2014); \textit{Oracle Am., Inc. v. Google Inc.}, 798 F. Supp. 2d 1111 N.D. Cal. 2011; \textit{Suffolk Techs. LLC v. Aol Inc.}, No. 1:12cv625, 2013 U.S. Dist. LEXIS 64630 (E.D. Va. Apr. 12, 2013); \textit{Limelight Networks, Inc. v. Xo Communs., LLC} Civil Action No. 3:15-CV-720-JAG, 2018 U.S. Dist. LEXIS 17802 (E.D. Va. Feb. 2, 2018).}. This has caused judges to criticize the NBS solution when determining a reasonable royalty \citep{wyatt2014keeping,sidak2015bargaining, jarosz2012hypothetical, yang2014damaging}.  Because Nash's solution is often not tailored to the specific facts of the case, parties are not taking full advantage of guideline fifteen. Another reason for criticism is the NBS is not simple to calculate or easy to interpret so that a court or jury can easily apply it \citep{wyatt2014keeping}.\footnote{\cite{wyatt2014keeping} gives two reasons why courts are reluctant to use the NBS: "First, damages experts often use the NBS improperly, failing to apply the specific facts of the case to their calculations [internal citation omitted]. Second, damages experts typically fail to adequately explain the NBS to courts and juries [internal citation omitted]."} To demystify the NBS, certain normalizations are introduced that provide for a simple calculation of damages. These normalizations make the NBS a powerful tool to value intellectual property and provide guidance in assigning proper compensation.

First, this paper applies Nash's solution in a more business-friendly manner by using terminology common on financial statements. Additionally, this paper normalizes each term in the NBS by the operating income. By doing this, the parties can better interpret the NBS and do not need to know exact dollar amounts when determining a royalty. The Choi and Weinstein \citep{choi2001analytical} Two Supplier World (TSW) model is the basis for our modifications.

Second, Nash's original solution assigns equal bargaining strength to each party. However, this equal bargaining strength assumption is, in general, not realistic \citep{higgins2015application}. This paper shows Nash's solution with an arbitrary bargaining weight to account for unequal bargaining strengths and presents a methodology for determining those strengths.

Third, a nomograph of the NBS is supplied to make it easy for parties to obtain a reasonable royalty using a simple straight edge graphically. Nomographs are useful to provide visualization so the NBS can be better explained.

By taking these steps, parties can take advantage of \textit{Georgia-Pacific} factor fifteen by allowing the NBS to be tailored to the specific facts of the case. This paper attempts to bring clarity to the use of the NBS, so the royalty that is assigned is both legally defensible and mutually beneficial.


\section{Elements of a Licensing Bargain}

The NBS is recast into a simple normalized form using common terms found on a financial statement to introduce common business terminology. 
 
\subsection{Operating Revenue} The operating revenue is the revenue generated from the intellectual property and is denoted by $O_R$. It does not include income from unusual events or income that is not primarily due to the use of the intellectual property.

\subsection{Operating Cost} The operating cost is the expense associated with producing and selling the product incorporating the intellectual property. This is defined as $O_C$ and does not include expenses from non-primary sources or unusual events. 

\subsection{Operating Income} The operating income, or profit, is determined by subtracting the operating cost from the operating revenue:  $O_I= O_R-O_C$. In formulating the asymmetric NBS, the licensor and licensee's operating income are denoted by $\pi_1$ and $\pi_2$, respectively, where the total profit in the system is $O_I$.

\subsection{Operating Margin}
The operating margin, $O_M$, is operating income divided by operating revenue and is expressed as $O_M = O_I/O_R$.

\subsection{Royalty}
The royalty is what the licensee will pay the licensor for the use of the intellectual property. There are two common ways to calculate a royalty. One way is to assign a royalty on each unit sold. The other is obtaining a royalty based on a percentage of revenue \citep{goldscheider2006licensing} by multiplying the revenue with the royalty rate, $r$.  In this paper, the focus is solely on a royalty based on revenue.

\subsection{Disagreement Payoffs}

A disagreement payoff is the opportunity cost of making the deal. In other words, disagreement payoffs are profits that come from a hypothetical negotiation that did not happen but could have happened if the parties did not agree to a deal. Disagreement payoffs are typically expressed as monetary amounts and are represented in this paper by $d_1$ and $d_2$ for the licensor and licensee, respectively. However, for computational ease, the disagreement payoffs are normalized by the operating income, and these are expressed as $d_1^\dagger$ and $d_2^\dagger$ for the licensor and licensee, respectively. A normalized disagreement payoff equal to one implies a party is indifferent between making the deal and not making the deal since the party could earn the same profit regardless.  For emphasis, a normalized disagreement payoff of $d_2^\dagger=0.5$ means the licensee's opportunity cost is half the total profit that a deal with the licensor can generate. Each parties' normalized disagreement payoffs can vary between zero and one. However, the sum of the normalized disagreement payoffs cannot exceed one, or a deal cannot be made since there is not enough profit to give each party their opportunity cost. The disagreement point is denoted by $d^\dagger = \left(d_1^\dagger,d_2^\dagger\right)$.

\subsection{Bargaining Weight}

A bargaining weight quantifies each party's influence in the negotiation and determines how the parties split the surplus from making the deal. The licensor's bargaining weight is $\alpha$, and the bargaining weight for the licensee is 1-$\alpha$, where the weight is between zero and one. The bargaining weight encapsulates how each party perceives their own and each other's bargaining strengths in a negotiation. The larger a party's bargaining weight, the more influence that party has in the negotiation. This means the party with the larger weight will obtain more surplus from making the deal. When applying the NBS, it has been common practice to assign each party a weight equal to 1/2, which implies that each party has the same influence in the negotiation \citep{ jarosz2012hypothetical,lemley2006patent,Jonathan2004BargainingAT}.


\section{The Asymmetric Nash Bargaining Solution}

John Nash developed the NBS, which provides a method for two parties who enter a profit-making agreement to determine how to share those profits optimally\citep{nash1950bargaining,nash1953two}. The axioms that satisfy the classic NBS are:

\begin{enumerate}
\item{\textbf{Individual rationality:}  No party will agree to accept a payoff lower than the one guaranteed to him under disagreement.}
\item{\textbf{Pareto optimality:} None of the parties can be made better off
without making at least one party worse off.}
\item{\textbf{Symmetry:} If the parties are indistinguishable, the agreement should not discriminate between them.}
\item{\textbf{Affine transformation invariance:} An affine transformation of the payoff and disagreement point should not alter the outcome of the bargaining process.}
\item{\textbf{Independence of irrelevant alternatives:} All threats the parties might make have been accounted  for in the disagreement point.}

\end{enumerate}

However, the introduction of a bargaining weight into the NBS allows the parties to be distinguishable when $d_1^\dagger=d_2^\dagger$ (potentially violating symmetry), known as the asymmetric NBS \citep{muthoo1999bargaining}. An excellent summary of the literature involving the asymmetric NBS and its use in intellectual property litigation is found in Bhattacharya\citep{bhattacharya2019nash}.  The bargaining weight can be influenced by other forces or tactics employed by the parties, which can be independent of the disagreement payoffs. These forces should be accounted for because they ultimately affect how the surplus is divided\footnote{ \cite{muthoo1999bargaining} states:  "However, the outcome of a bargaining situation may be influenced by other forces (or, variables), such as the tactics employed by the bargainers, the procedure through which negotiations are conducted, the information structure and the players' discount rates. However, none of these forces seem to affect the two objects upon which the NBS is defined [the disagreement payoffs], and yet it seems reasonable not to rule out the possibility that such forces may have a significant impact on the bargaining outcome." }.

The asymmetric NBS is formed from the constrained maximization problem\citep{kalai1977nonsymmetric,binmore1986nash,roth2012axiomatic}:

\begin{equation}
\label{eq:NBTS1}
\max_{\pi_1 , \pi_2} \left( \pi_1 - d_1\right)^{\alpha} \left(\pi_2 - d_2\right)^{1-\alpha}
\end{equation}

\noindent
Subject to the following conditions:
\begin{equation}
\label{eq:NBTS2}
\pi_1 \geq d_1
\end{equation}

\begin{equation}
\label{eq:NBTS3}
\pi_2 \geq d_2
\end{equation}

\begin{equation}
\label{eq:NBTS4}
\pi_1 + \pi_2 \leq O_I
\end{equation}

\noindent
Maximum occurs when:
\begin{equation}
\label{eq:NBTS5}
(1-\alpha) \, \left(\pi_1^* - d_1\right) = \alpha \, \left(\pi_2^* - d_2\right)
\end{equation}

\noindent

\begin{equation}
\label{eq:NBTS6}
\pi_1^* + \pi_2^* = O_I
\end{equation}

\noindent

Solving for the optimal partition of the profits gives the final result:

\begin{subequations}\label{eq:NBTS7}
\begin{align}
\pi_1^* &= d_1 + \alpha \left( O_I - d_1 -d_2 \right) \label{sub-eq-1:1} \\ 
\pi_2^* &= d_2+ (1-\alpha) \left( O_I - d_1 -d_2 \right) \label{sub-eq-2:1} 
\end{align}
\end{subequations}

Eq. \eqref{eq:NBTS7}'s interpretation is that the parties first agree to give each other their respective disagreement payoffs and split the remaining profit (surplus) according to their bargaining strength.

\subsection{Normalized Royalty Model}

To make the TSW model more practical, Eq. \eqref{eq:NBTS7} was modified to introduce a royalty based on a percentage of revenue. Moreover, by simple algebraic manipulation,  Eq. \eqref{eq:NBTS7} can be modified where every term is normalized by the operating income and varies between zero and one. Having each term normalized is powerful because the parties do not need to think about specific dollar amounts. Instead, the parties can think in terms of fractions of profit.


The licensor is referred to as party 1 and the licensee as party 2. Under these assumptions, the payoffs for party 1 and 2 are:

\noindent

\begin{equation}
\label{eq:NBTS8}
\frac{\pi_1^*}{O_I} =  \frac{r  \, O_R}{O_I} = \frac{r}{O_M}
\end{equation}

\noindent
\begin{equation}
\label{eq:NBTS12}
\frac{\pi_2^*}{O_I} =  \frac{O_R-O_C-rO_R}{O_I} = 1-\frac{r}{O_M}
\end{equation}


\noindent
Additionally defining:

\begin{equation}
\label{eq:NBTS9}
d_1^\dagger = \frac{d_1}{O_I}  ~~~~~ 0 \leq d_1^\dagger \leq 1
\end{equation}

\noindent

\begin{equation}
\label{eq:NBTS10}
d_2^\dagger = \frac{d_2} {O_I}   ~~~~~ 0 \leq d_2^\dagger \leq 1
\end{equation}

\noindent

Substituting Eqs. \eqref{eq:NBTS8}, and \eqref{eq:NBTS9}--\eqref{eq:NBTS10} into Eq. \eqref{sub-eq-1:1}, the result for the optimal NBS is obtained with an arbitrary bargaining weight for party 1:

\noindent

\begin{equation}
\label{eq:NBTS13}
\frac{r}{O_M}= d_1^\dagger + \alpha  \left(1 - d_1^\dagger - d_2^\dagger \right)
\end{equation}

\noindent
Where:
\begin{equation}
\label{eq:NBTS14}
0 \leq d_1^\dagger + d_2^\dagger \leq 1
\end{equation}

To maintain Pareto efficiency, Eq. \eqref{eq:NBTS13} must satisfy the following \citep{lee1980theory}:

\begin{subequations}\label{eq:NBTS15}
\begin{align}
\frac{\partial r}{\partial d_1^\dagger} &>  0 \label{sub-eq-1:2} \\
\frac{\partial r}{\partial d_2^\dagger} &< 0 \label{sub-eq-2:2}
\end{align}
\end{subequations}

The interpretation of Eq. \eqref{eq:NBTS15} is that for a small positive change in party 1's disagreement payoff, the royalty should increase.  In contrast, for a small positive change in party 2's disagreement payoff, the royalty should decrease - that is, a party cannot be made better off without making the other party worse off.


\section{Estimation of the Bargaining Weight}

The bargaining weight, $\alpha$, represents how the parties perceive their bargaining strength and how they see the other's bargaining strength. To account for all the perceptions of bargaining strength, the parameter, $P_{m,n}$, is introduced as party m's bargaining strength as perceived by party n. For example, $P_{1,2}$ is how the licensee perceives the licensor's bargaining strength. 

Making the simple assumption that the bargaining strength of each party is the average of their perception and the perception of the other party, the following mathematical ansatz is introduced using two different equations to describe the bargaining weight of party 1:


\begin{subequations}\label{eq:NBTS17}
\begin{align}
\alpha_1 &= \frac{1}{2}\left[P_{1,1} + P_{1,2}\right] \label{sub-eq-1:3} \\
\alpha_2 &= 1-\frac{1}{2}\left[P_{2,1} + P_{2,2}\right] \label{sub-eq-2:3}
\end{align}
\end{subequations}
Averaging Eqs. \eqref{sub-eq-1:3}--\eqref{sub-eq-2:3}, the complete expression for the bargaining weight of party 1 is obtained:
 
\begin{align}
\label{eq:NBTS18}
\alpha &\equiv \frac{1}{2} \left[\alpha_1 + \alpha_2\right] \nonumber \\
    &=\frac{1}{2} + \frac{1}{4}\left[P_{1,1} + P_{1,2} -P_{2,1} - P_{2,2}\right] ~~~~  0 \leq P_{m,n} \leq 1
\end{align}

Eq. \eqref{eq:NBTS18} is critically important because a simple procedure now exists to define the bargaining weight of party 1. By formally defining the bargaining weight, each party's bargaining strengths can be incorporated to fit the particular facts of a case.

There are three basic approaches when calculating a bargaining weight. One approach is to treat $\alpha$ as a function that is independent of the disagreement payoffs. The second approach is to make the bargaining weight strictly a function of the disagreement payoffs. The third is a mixture of the first two approaches.


\subsection{The Classic Nash Bargaining Solution}

When $P_{1,1} + P_{1,2} = P_{2,1} + P_{2,2}$ in Eq. \eqref{eq:NBTS18}, then  $\alpha=1/2$ and the classic symmetric NBS is obtained:

\begin{equation}
\label{eq:NBTS19}
\frac{r}{O_M}=\frac{1}{2}\left(1 + d_1^\dagger - d_2^\dagger\right)
\end{equation}

\begin{figure}[]
\begin{center}
\includegraphics[scale=.27]{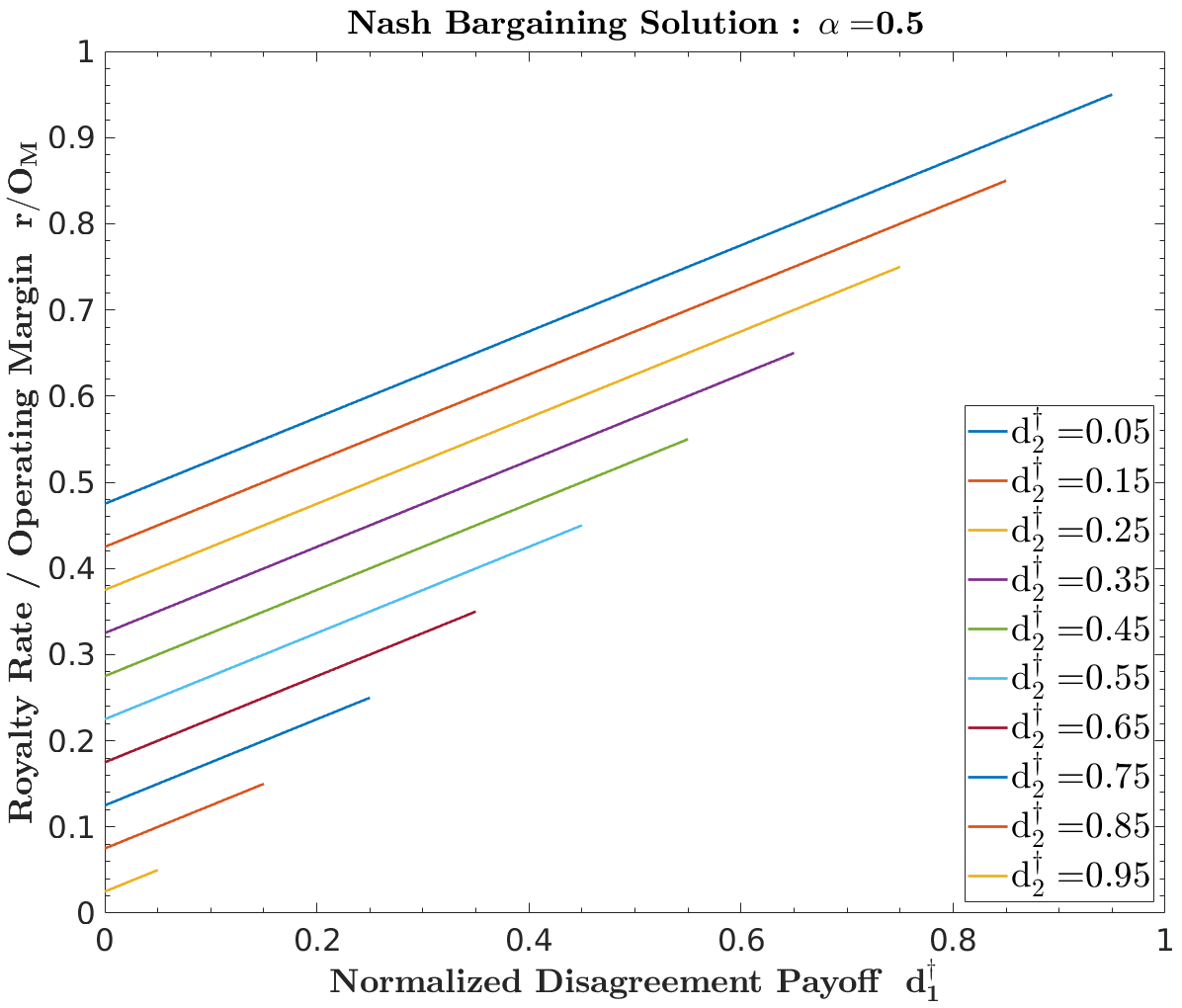} 
\end{center}
\caption{Family of Nash Bargaining Solutions Given Equal Bargaining Power}
\label{fig:ADI_Nash1}
\end{figure}

Fig. \ref{fig:ADI_Nash1} presents the family of solutions of Eq. \eqref{eq:NBTS19}.  Note that the lines of equal $d_2^\dagger$ are linear and equidistant from each other. Also, note that the lines are not the same length due to the constraint of Eq. \eqref{eq:NBTS14}.


\section{Discussion}
In this section, some hypothetical situations are presented to demonstrate the use of the NBS.  Since the assignment of a party's perception of bargaining strength to a particular $P_{m,n}$ can be somewhat arbitrary, the examples given in this section are for illustration only. In the end, it is the job of the parties to provide a careful assessment of each of their perceptions and incorporate them properly into Eq. \eqref{eq:NBTS18}. By choosing these perceptions, the NBS can be applied to the specific facts of the case.

\subsection{Estimation of Bargaining Strengths Independent of the Disagreement Payoffs }

The use of Eq. \eqref{eq:NBTS18} is demonstrated by a simple hypothetical negotiation involving bargaining strengths independent of the disagreement payoffs.

\subsubsection{Number of Competitors as Strength} 
\label{sec:competitors}

The bargaining strength of party 1 is dependent on the relation between the hypothetical number of licensors and licensees in the market \citep{zimmeck2011game}. This is because if party 1 has a wide range of options to sell its intellectual property, then party 1 is presumably less worried about making a deal with party 2. After all, the licensor can credibly walk away and license the technology to another firm. Therefore, if party 1 can sell its intellectual property to multiple licensees, the expectation is that party 1 has more bargaining strength. Conversely, if party 2 can license an acceptable substitute, party 1's bargaining strength will diminish.  The following equation is driven by the ratio of the number of licensors to the number of licensees in the relevant market \citep{zimmeck2011game}. The component of party 1's bargaining strength as derived from the number of licensors and licensees in the market is:

\begin{equation}
\label{eq:NBTS20}
P_{1,n}^L=1-\mathrm{min}\left[1,\frac{ \mathrm{Licensors}}{\mathrm{Licensees}}\right] 
\end{equation}

The perception is assigned as $P_{1,n}$ because either party may perceive \eqref{eq:NBTS20} as a component of party 1's bargaining strength.



\subsubsection{Market Share as Strength}
\label{sec:market}
In business, market share is regarded as the essential element of dominance \citep{lee2019industrial}. As a result, valuing a component of party 1's bargaining strength by the amount of market share, $s$, is attractive instead of a measurement of potential profits. Using potential profits as a measurement of bargaining strength may not be appealing because profits are highly variable from year to year while market share is relatively constant over long periods of time. Additionally, courts often measure a firm's dominance by market share rather than profits \citep{10.2307/2487731}. Therefore, another measurement of bargaining strength is determining how much market share party 2 would gain as a result of the deal.  The component of party 1's bargaining strength, as derived from market share, is:

\begin{equation}
\label{eq:NBTS21}
P_{1,n}^S=\frac{s}{S}  ~~~~~~~~~~~~~~~~~~~ 0\leq s\leq S
\end{equation}

In Eq. \eqref{eq:NBTS21}, $S$ denotes that fraction of the total market party 2 realistically desires.

\subsubsection{Life of the Patent as Strength} 
\label{sec:patent}
Another perception of strength can be the time left until the patent expires.  Presumably, party 1 is in a strong bargaining position when the patent is recently issued but is in a weak bargaining position when the patent is about to expire. Let the patent's life be denoted by $T$ and the time elapsed since issue by $t$.  The component of party 1's bargaining strength as derived from patent life is:

\begin{equation}
\label{eq:NBTS22}
P_{1,n}^T= 1-\frac{t}{T} ~~~~~~~~~~~~~~~  0\leq t \leq T
\end{equation}

\subsubsection{Example} 
\label{sec:constant}

In this hypothetical example, party 1 perceives its bargaining strengths, with equal weight, the lack of acceptable substitutes for its patent, and the potential market share that the patent can bring to party 2.  Party 2 perceives party 1's bargaining strength as only the life of the patent.  Party 2 has a unique manufacturing base that can take full advantage of party 1's patent and perceives its bargaining strength as $P_{2,2}=2/3$.  Party 1 is aware of party 2's unique manufacturing capabilities but only perceives party 2's strength as $P_{2,1} = 1/2$. 

Substituting each perception into Eq. \eqref{eq:NBTS18}: 

\begin{equation}
\label{eq:NBTS23}
\alpha= \frac{1}{2} + \frac{1}{4}\left[\frac{P_{1,1}^L+P_{1,1}^S}{2} + P_{1,2}^T -\frac{1}{2} - \frac{2}{3}\right]
\end{equation}

Eq. \eqref{eq:NBTS23} can now be substituted into Eq. \eqref{eq:NBTS13} to obtain the royalty for party 1.

\subsection{Estimation of Bargaining Strengths Using Disagreement Payoffs }

Disagreement payoffs can be a reasonable measure of bargaining strength because the parties can potentially walk away from the negotiation based on the disagreement payoffs alone. Therefore, $\alpha$ can be a function of each party's disagreement payoff. This approach requires the least amount of information but requires the parties to determine a functional form of $\alpha\left(d_1^\dagger,d_2^\dagger\right)$ that adequately represents the negotiation. For a standard of fairness, it is stipulated that when $d_1^\dagger= d_2^\dagger$, the parties should split the surplus equally, which implies that symmetry is reintroduced. It is possible to construct an $\alpha\left(d_1^\dagger,d_2^\dagger\right)$ that reintroduces symmetry and yet provides variability in the bargaining weight. 

Cases 1-3 in Table \ref{tab:symmetric} are examples of symmetric bargaining weights driven by the  parties' disagreement payoffs.

\begin{table*}[]
\caption{Three Cases of Symmetric Disagreement Payoff Driven Bargaining Weights}
\label{tab:symmetric}
\begin{tabular}{|c|c|c|c|c|c| }
\hline
\textbf{Case} & \textbf{P\textsubscript{1,1}} & \textbf{P\textsubscript{1,2}} & \textbf{P\textsubscript{2,1}} & \textbf{P\textsubscript{2,2}} & $\mathbf{\alpha}\left(d_1^\dagger,d_2^\dagger\right)$ \\ 
\hline
\textbf{1} & $d_1^\dagger$ & $d_1^\dagger$ & $ d_2^\dagger$ & $d_2^\dagger$ & $\frac{1}{2} + \frac{d_1^\dagger - d_2^\dagger}{2}$ \\
\hline
\textbf{2} & $\frac{d_1^\dagger}{d_1^\dagger + d_2^\dagger}$ & $\frac{d_1^\dagger}{d_1^\dagger+ d_2^\dagger}$ & $\frac{d_2^\dagger}{d_1^\dagger+ d_2^\dagger}$ & $\frac{d_2^\dagger}{d_1^\dagger + d_2^\dagger}$ &  $\frac{d_1^\dagger}{d_1^\dagger + d_2^\dagger}$\\
\hline
\textbf{3} & $\frac{d_1^\dagger}{d_1^\dagger + d_2^\dagger}$ & $\frac{d_1^\dagger}{d_1^\dagger + d_2^\dagger}$  & $\frac{1-d_1^\dagger}{2-d_1^\dagger-d_2^\dagger}$ & $\frac{1-d_1^\dagger}{2-d_1^\dagger-d_2^\dagger}$ & ${\frac {{{ d_1^\dagger}}^{2}+ \left( 2\,{ d_2^\dagger}-3 \right) { d_1^\dagger}+{{d_2^\dagger}}^{2}-{d_2^\dagger}}{ 2 \, \left( {d_1^\dagger}+{d_2^\dagger} \right)  \left( -2+{d_1^\dagger}+{d_2^\dagger} \right) }}$ \\
\hline
\end{tabular}
\end{table*}



\subsubsection{Case 1}

In Case 1 of Table \ref{tab:symmetric}, each party assumes that its bargaining strength is equal to its disagreement payoff.  Moreover, each party agrees that the other party's bargaining strength is its disagreement payoff. Substituting Case 1 of Table \ref{tab:symmetric}  into Eq. \eqref{eq:NBTS13}: 

\begin{equation}
\label{eq:NBTS24}
\frac{r}{O_M}=  \frac{{d_2^\dagger}^2 - {d_1^\dagger}^2 + 2\left(d_1^\dagger-d_2^\dagger\right) + 1}{2}
\end{equation}

\begin{figure}[]
\begin{center}
\includegraphics[scale=.27]{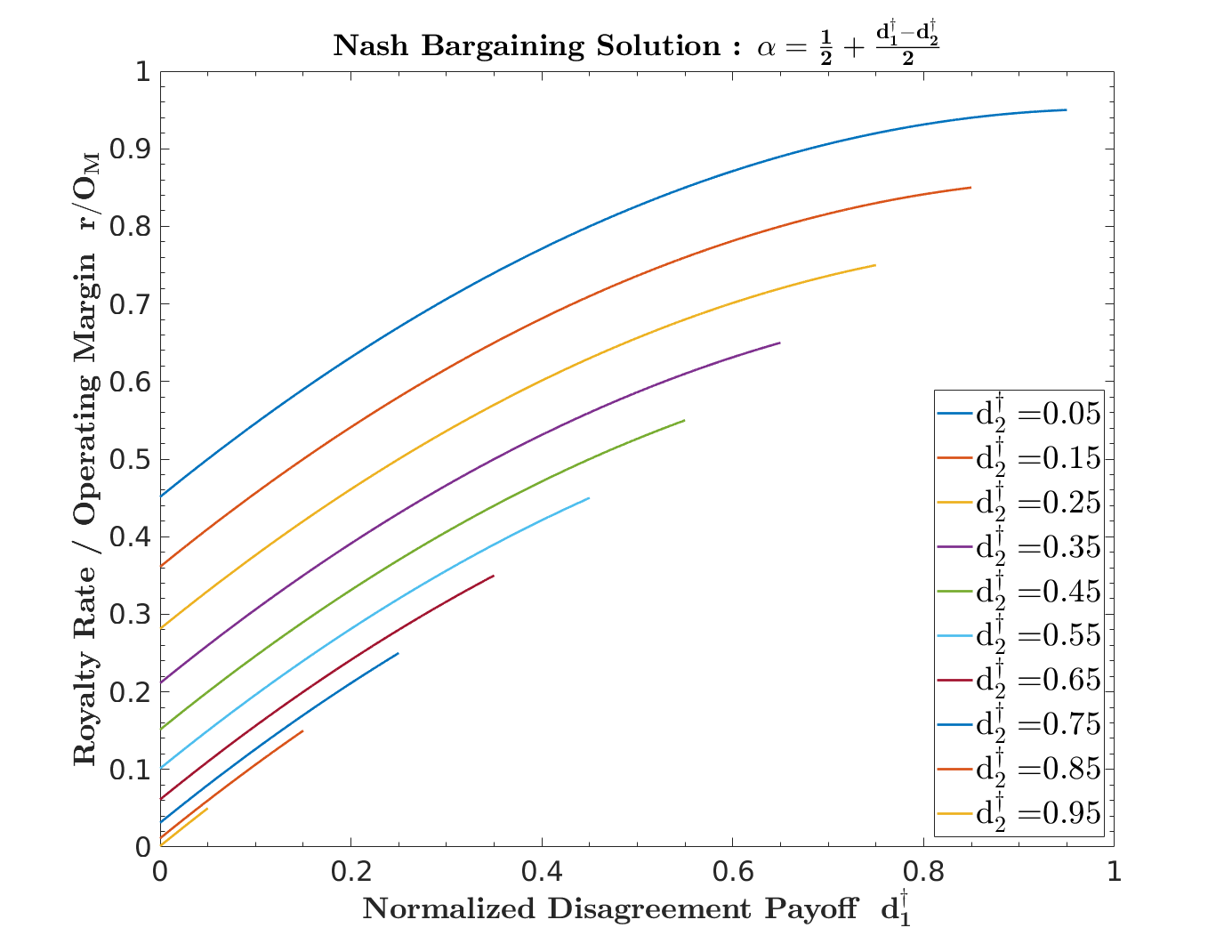} 
\end{center}
\caption{Family of Nash Bargaining Solutions for Table \ref{tab:symmetric} Case 1}
\label{fig:ADI_Nash2}
\end{figure}

Eq. \eqref{eq:NBTS24} shows a quadratic dependence on both $d_1^\dagger$ and $d_2^\dagger$, and this dependence is illustrated in Fig. \ref{fig:ADI_Nash2}.  Note that a party is penalized to a much greater extent for having a weak disagreement payoff position over the classic NBS of Fig. \ref{fig:ADI_Nash1}.

\subsubsection{Case 2}

In Case 2 of Table \ref{tab:symmetric}, each party assumes that its bargaining strength is equal to its fraction of the total disagreement payoff position $d_1^\dagger + d_2^\dagger$.  Moreover, each party agrees that the other party's bargaining strength is its fraction of the total disagreement payoff. Substituting Case 2 of Table \ref{tab:symmetric}  into Eq. \eqref{eq:NBTS13}:

\begin{equation}
\label{eq:NBTS25}
\frac{r}{O_M}=\frac{d_1^\dagger}{d_1^\dagger+d_2^\dagger}
\end{equation}

Interestingly, the payoff for each party is the party's bargaining weight.  Moreover, the solution is independent of $O_I$, which makes this a non-cooperative bargain and is equivalent to a limiting case of the Rubinstein model\cite{rubinstein1982perfect,binmore1986nash}\footnote{In \cite{muthoo1999bargaining}, the Subgame Perfect Equilibrium solution, where the time limit between offers $\Delta \rightarrow 0$, is presented in terms of discount rates ($r_A,r_B$) where $d_1^\dagger/d_2^\dagger = r_B/r_A$.  The payoff pair obtained through perpetual disagreement, the Impasse Point, is $\left(\mathcal{I}_A,\mathcal{I}_B\right)=\left(d_1^\dagger,d_2^\dagger\right)$. See Corollary 3.1 and Definition 3.1.}, where the parties take turns in making an offer until an agreement is secured.\footnote{\cite{muthoo1999bargaining} discusses the Rubinstein model, where the parties take turns in making an offer until an agreement is secured. "...Another insight is that a party's bargaining power depends on the relative magnitude of the parties' respective costs of haggling, with the absolute magnitudes of these costs being irrelevant to the bargaining outcome. ...In a boxing match, the winner is the relatively stronger of the two boxers; the absolute strengths of the boxers are irrelevant to the outcome."}

\begin{figure}[]
\begin{center}
\includegraphics[scale=.27]{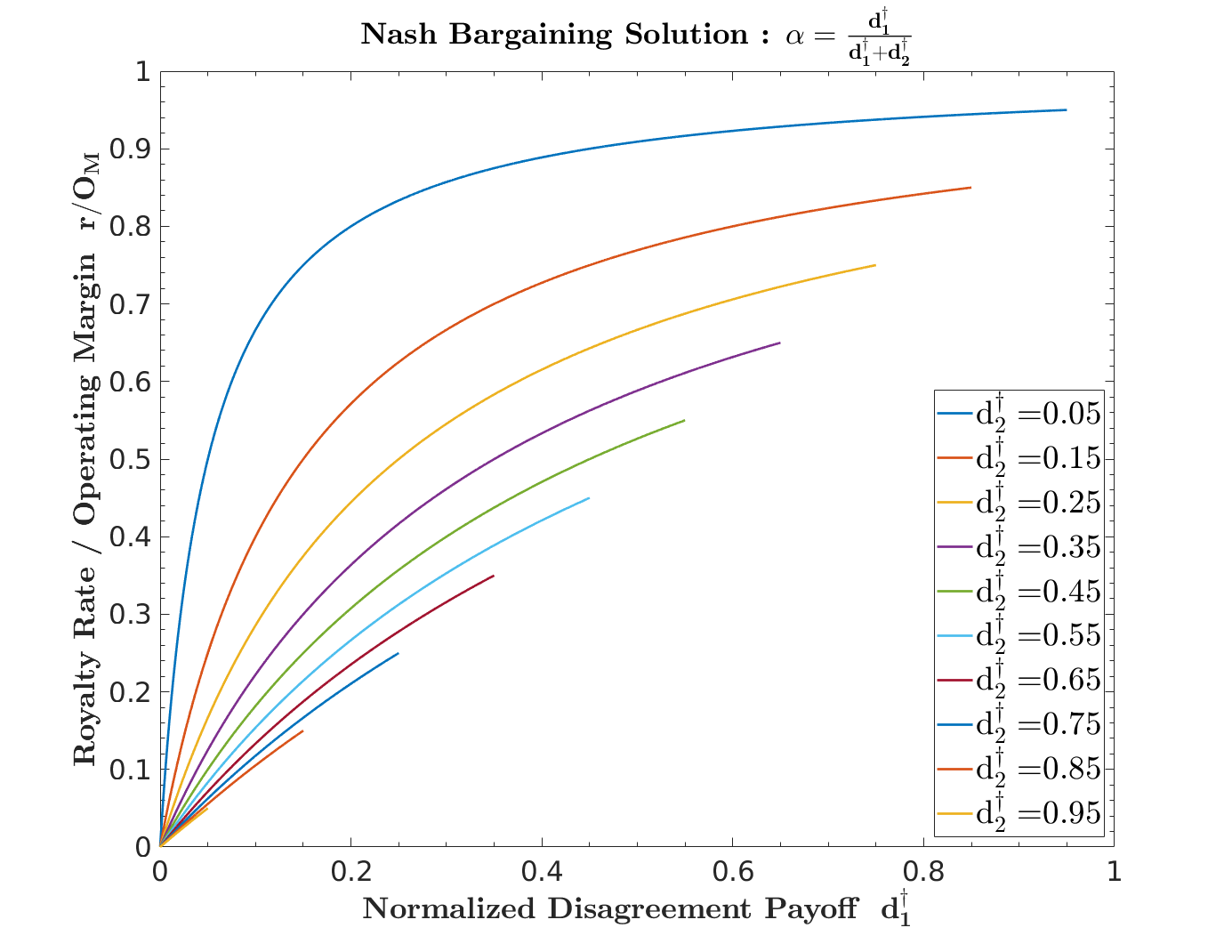} 
\end{center}
\caption{Family of Nash Bargaining Solutions for Table \ref{tab:symmetric} Case 2}
\label{fig:ADI_Nash3}
\end{figure}

Fig. \ref{fig:ADI_Nash3} shows the family of solutions for Eq. \eqref{eq:NBTS25}.  Note the rapid collapse to zero of party 1's royalty for any constant $d_2^\dagger$ as $d_1^\dagger$ approaches zero\footnote{In antitrust litigation, Case 1 or Case 2 could be used to set a threshold on the bargaining weight where one firm is shown to have significantly more bargaining power to trigger litigation. For example, if $\alpha \geq 0.75$ in Case 1, this could be a threshold for which litigation may be warranted. A notable antitrust case that uses the NBS is \textit{United States v. AT\&T, Inc.}, 310 F. Supp. 3d 161, 164 (D.D.C. 2018), aff'd, 916 F.3d 1029 (D.C. Cir. 2019).}.


\subsubsection{Case 3}

Case 3 presents an example where party 2's bargaining strength depends on party 1's weakness.  As in the previous examples, all parties agree on each other's bargaining strength.   Substituting Case 3 of Table \ref{tab:symmetric} into Eq. \eqref{eq:NBTS13}: 

\begin{equation}
\label{eq:NBTS26}
\frac{r}{O_M}=\frac{{d_2^\dagger}^2-{d_1^\dagger}^2-2d_2^\dagger+d_1^\dagger+1}{2-d_1^\dagger-d_2^\dagger}
\end{equation}

\begin{figure}[]
\begin{center}
\includegraphics[scale=.27]{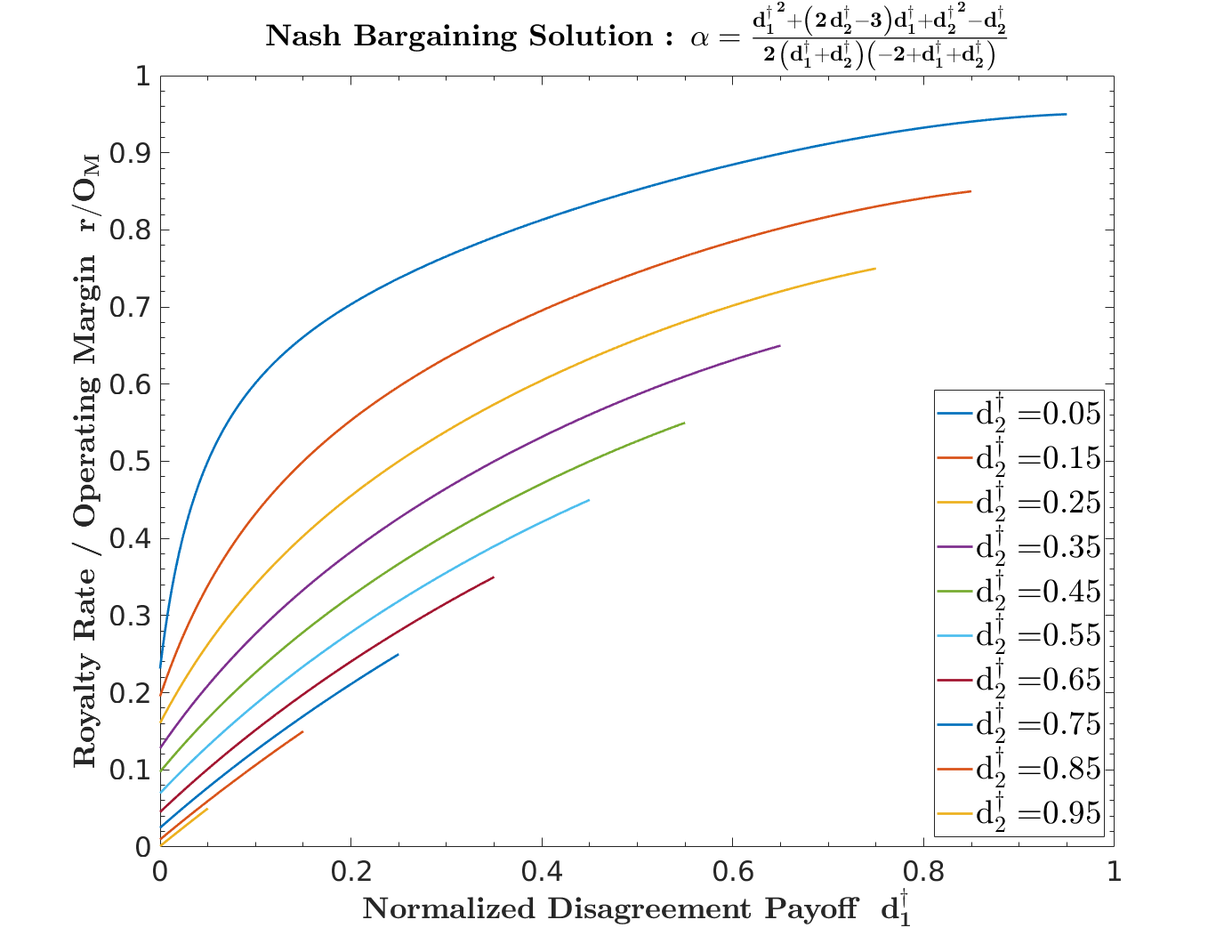} 
\end{center}
\caption{Family of Nash Bargaining Solutions for Table \ref{tab:symmetric} Case 3}
\label{fig:ADI_Nash4}
\end{figure}

Fig. \ref{fig:ADI_Nash4} shows the family of solutions for Eq. \eqref{eq:NBTS26}. The figure shows the same quadratic dependence as Case 1 Fig. \ref{fig:ADI_Nash2}, where the lines of constant $d_2^\dagger$ get closer together as $d_2^\dagger$  becomes dominant.  Party 1's bargaining advantage has increased from Case 2 for small $d_1^\dagger$ because party 2's strength is derived from party 1's weakness and not its strength as in Case 2.

\subsection{Estimation of Bargaining Strength Using Combinations}
Perceptions, independent or dependent of the disagreement payoffs can be combined in Eq. \eqref{eq:NBTS18}. However, there are cases when combinations of perceptions are not Pareto efficient, and this is examined next.

\subsubsection{Solutions That Violate Pareto Efficiency} 

\begin{figure}[]
\begin{center}
\includegraphics[scale=.27]{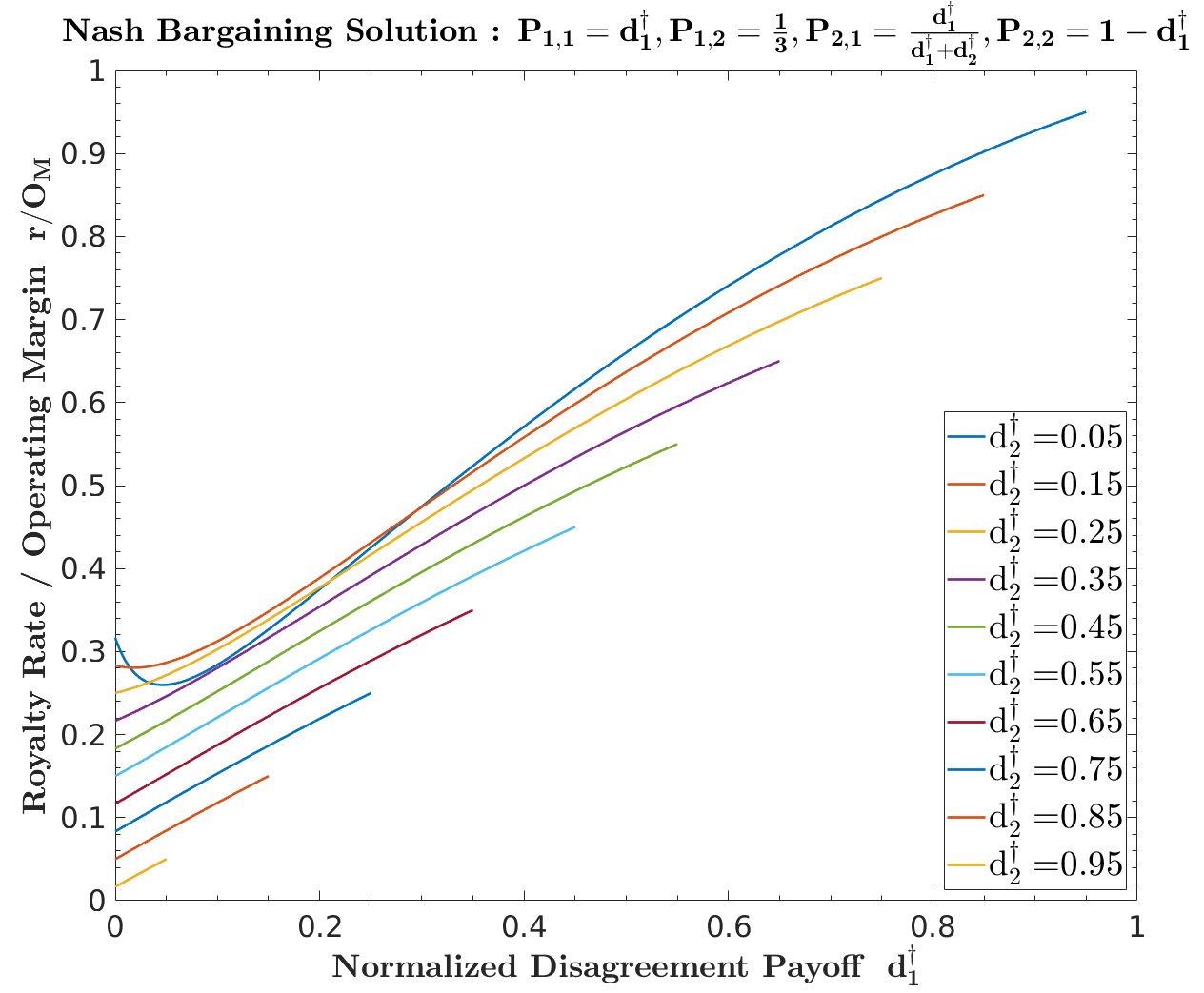} 
\end{center}
\caption{Family of Nash Bargaining Solutions With Regions That Violate Pareto Efficiency}
\label{fig:ADI_Nash6}
\end{figure}

When $\alpha$ is a function of the disagreement payoffs, there can be combinations of perceptions that violate Pareto efficiency in a part of the solution space. Fig. \ref{fig:ADI_Nash6} is one such example.
Substituting the following hypothetical $\alpha$ into Eq. \eqref{eq:NBTS13}, Fig. \ref{fig:ADI_Nash6} is obtained:

\begin{equation}
\label{eq:NBTS27}
\alpha\left(d_1^\dagger,d_2^\dagger\right)=\frac{1}{2} + \frac{1}{4}\left[d_1^\dagger + \frac{1}{3} -\frac{d_1^\dagger}{d_1^\dagger +d_2^\dagger} - \left(1-d_1^\dagger\right)\right]
\end{equation}

From Fig. \ref{fig:ADI_Nash6}, it can be seen that the solution space is not Pareto efficient everywhere because when both $d_1^\dagger$ and $d_2^\dagger$ are small, party 1 will receive a lower royalty for a slight increase in $d_1^\dagger$, which is counterintuitive. 

It is easily shown that the royalty in Fig. \ref{fig:ADI_Nash6} violates Eq. \eqref{eq:NBTS15} when $d_2^\dagger$ is small. The reason for this violation is that the specification of $P_{2,1}$ causes party 2's strength as perceived  by party 1 to be lower as party 1's disagreement payoff lowers. This influences a small section of the solution space to violate Pareto efficiency.  



\section{Nomographs}

\begin{figure}[]
\begin{center}
\includegraphics[scale=.16]{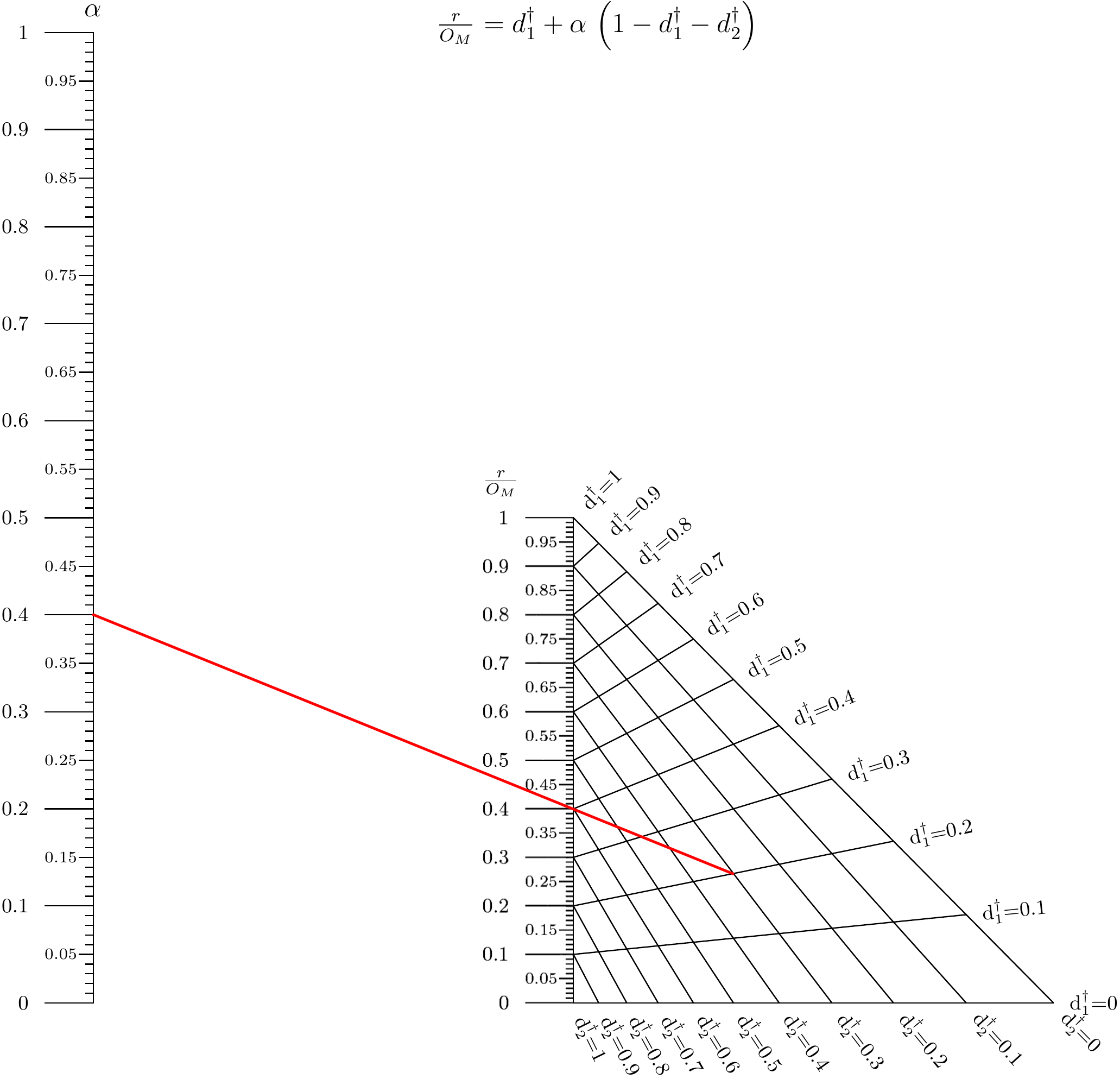} 
\end{center}
\caption{The use of the nomograph is demonstrated with  $d_1^\dagger = 0.20$, $d_2^\dagger = 0.30$, and $\alpha = 0.40$ to solve for $r/O_M$ = 0.40.}
\label{fig:ADI_Nash7}
\end{figure}

To make it easy to compute a royalty using the asymmetric NBS, a nomograph was constructed (see Fig. \ref{fig:ADI_Nash7}) with PyNomo \citep{PyNomo,Doerfler}.\footnote{ Type 9 General Determinant was used.} A nomograph is a diagram that is a graphical representation of a mathematical function. It allows for quick computation without the need to substitute numbers into a formula. Nomographs also provide visualization as to how the asymmetric NBS behaves so it can be easily explained.

To use the nomograph, pick any three variables on the graph and draw a straight line to get the fourth variable. For example, suppose that the normalized disagreement payoffs are $\mathrm{d}_1^\dagger=0.20$ and $\mathrm{d}_2^\dagger=0.30$. Additionally, suppose $\alpha=0.40$. Using a straight edge, a line is drawn from  $\alpha=0.40$ to a point on the grid where  $(\mathrm{d}_1^\dagger,\mathrm{d}_2^\dagger) = (0.20,0.30)$.  The royalty for party 1 is read off the corresponding scale.

A blank nomograph is provided in the appendix.


\section{Conclusion}

In this model of the asymmetric NBS, there are three essential variables needed to obtain a royalty. They are the disagreement payoffs of both party 1 and party 2, and the bargaining weight. At a minimum, the parties should have a good understanding of the licensed product's operating margin if a royalty rate is to be computed along with the need to make educated guesses on the disagreement payoffs of both parties. Various examples were given to demonstrate how each party's bargaining strengths can be incorporated into the bargaining weight. These individual bargaining strengths can be used to apply the NBS to the specific facts of the case. Although \textit{Georgia-Pacific} factor fifteen is the basis for this analysis, the other fourteen factors could also be used to obtain the normalized disagreement payoffs and choose the bargaining strengths. Finally, a nomograph has been produced so the parties can easily calculate the asymmetric NBS and solve for a reasonable royalty. 


%

\section*{Acknowledgments}
One of the authors (D.M. Kryskowski) would like to thank Professors Li Way Lee and Vitor Kamada of Wayne State University for their encouragement in pursuing this topic.  The author would also like to thank Professor  J.J. Prescott of the University of Michigan Law School for sparking the author's interest in Law \& Economics. 

\newpage
\bibliographystyle{model2-names}
\bibliography{IP}


\appendix 

\section{Blank Nomograph}
%

\begin{figure*}[]
\begin{center}
\includegraphics[scale=.29]{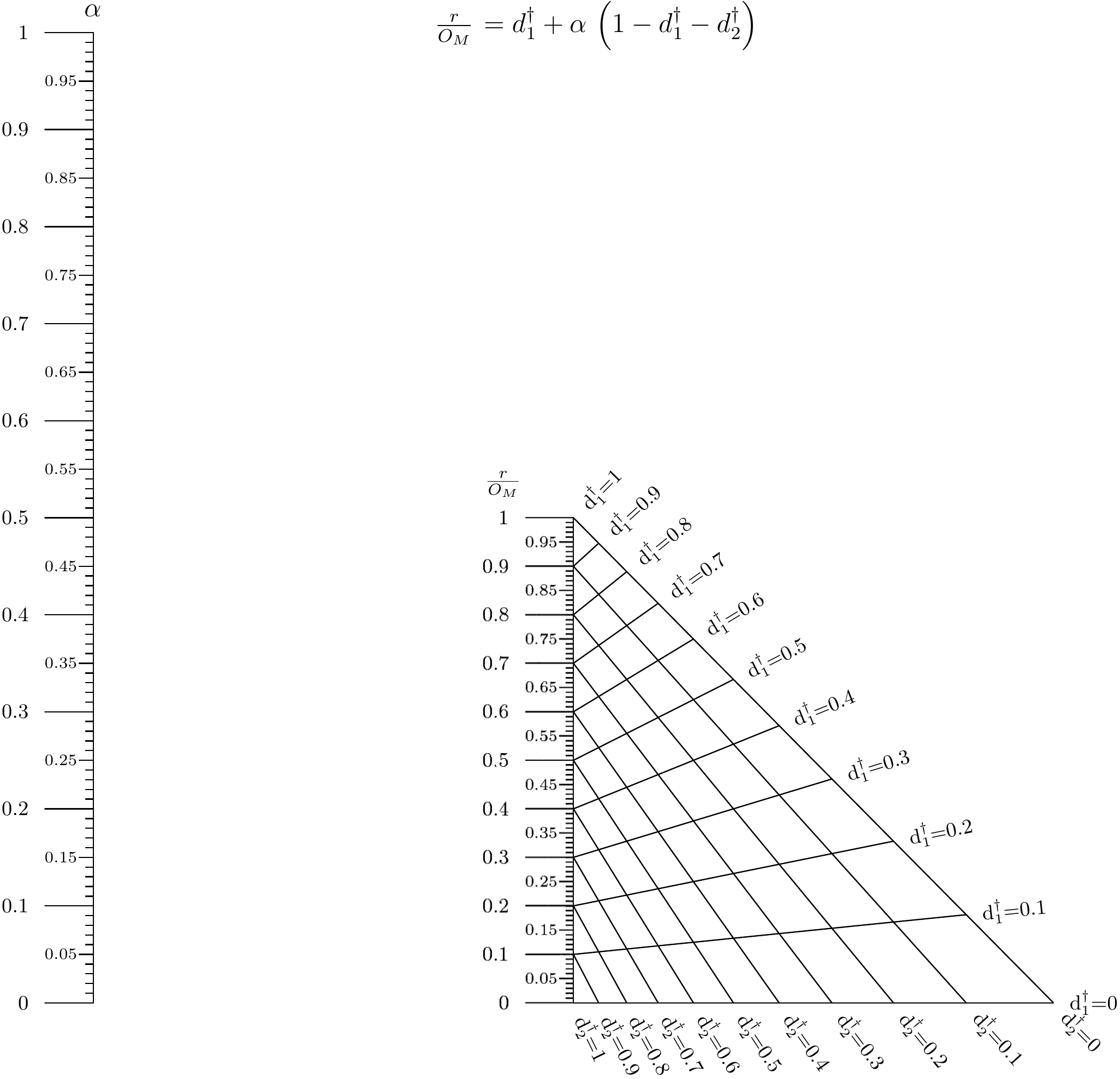} 
\end{center}
\caption{Blank Nomograph}
\label{fig:ADI_Nash8}
\end{figure*}

\end{document}